\documentclass[twocolumn,superscriptaddress,showpacs,oneside,prl]{revtex4}
\usepackage{amssymb}

\usepackage{epsfig,amssymb,amsfonts,amsmath}

\begin{document}

\title{Quantifying quantum discord and entanglement of formation via unified
purifications}
\author{Li-Xiang Cen}
\affiliation{Department of Physics, Sichuan University, Chengdu 610065,
China}\affiliation{Department of Chemistry, Hong Kong University of Science and
Technology, Kowloon, Hong Kong}

\author{Xin-Qi Li}
\affiliation{Department of Physics, Beijing Normal University, Beijing
100875, China}
\affiliation{Department of Physics, Sichuan University, Chengdu 610065,
China}\affiliation{Department of Chemistry, Hong Kong University of
Science and Technology, Kowloon, Hong Kong}

\author{Jiushu Shao}
\affiliation{College of Chemistry, Beijing Normal University, Beijing
100875, China}
\affiliation{Department of Chemistry, Hong Kong University of
Science and Technology, Kowloon, Hong Kong}

\author{YiJing Yan}
\affiliation{Department of Chemistry, Hong Kong University of Science and
Technology, Kowloon, Hong Kong}

\begin{abstract}
We propose a scheme to evaluate the amount of quantum discord and
entanglement of formation for mixed states, and reveal their ordering
relation via an intrinsic relationship between the two quantities
distributed in different partners of the associated purification. This
approach enables us to achieve analytical expressions of the two measures
for a sort of quantum states, such as an arbitrary two-qubit density matrix
reduced from pure three-qubit states and a class of rank-$2$ mixed states of
$4\times 2$ systems. Moreover, we apply the scheme to characterize fully the
dynamical behavior of quantum correlations for the specified physical
systems under decoherence.
\end{abstract}


\maketitle

 Quantum correlation constitutes a fundamental resource for quantum
information processing and it has been the subject of intensive studies in
the last decades. The non-locality aspect of quantum correlations, termed as
entanglement, was first singled out as the characteristic trait of quantum
mechanics that is inaccessible to classical objects \cite{schro}. It is
widely believed that entanglement constitutes the key ingredient leading to
the power of quantum computation \cite{shor,grove}. Operationally, entangled
states are those that cannot be prepared through local operations and
classical communication between two parties \cite{werner}. In other words,
they cannot be written as separable form: $\rho_{AB}\neq\sum_kp_k\rho_A^k
\otimes\rho_B^k$. The amount of entanglement of $\rho_{AB}$ shared in the
two parties is usually defined by \textit{entanglement of formation} (EoF)
\cite{bennett2}, i.e., the minimal average entanglement of pure state
ensembles to create $\rho _{AB}$: $E(\rho_{AB})\equiv\min\sum_kp_kE(|\psi_k
\rangle)$, where the minimum is taken over all possible decomposition $
\rho_{AB}=\sum_kp_k|\psi_k \rangle \langle \psi_k|$, and $E(|\psi_k \rangle)$
is entropy entanglement of the pure state $|\psi_k\rangle $ in the ensemble.

A different notion of measure, quantum discord $Q_{AB}$, has also been
proposed \cite{vedral,zurek} to characterize quantum correlations based on
an information-theoretic measure of mutual information. Distinctly, quantum
discord is figured out through quantifying the classicality and/or
nonclassicality in the total mutual information $I(\rho_{AB})\equiv
S(\rho_A)+S(\rho _B)-S(\rho _{AB})$. It is defined as $Q_{AB}\equiv I(\rho
_{AB})-J_{AB}$, the discrepancy between $I(\rho _{AB})$ and its classical
counterpart $J_{AB}\equiv S(\rho_B)-S(\rho_B|A)$, where $S(\rho_B|A)$ is the
conditional entropy, i.e., the minimal average entropy of $B$, given
measurements on $A$ [cf.~Eq.\;(\ref{condi})]. For pure bipartite states $
|\psi_{AB}\rangle $, discord is equal to the entropy entanglement, $
Q_{AB}=E(|\psi_{AB}\rangle )\equiv-\mathrm{{tr}(\rho_A\log\rho_A )}$, but
for mixed states the discord is generally not identical to EoF except for
particular cases. The conceptual difference of these two measures has
motivated extensive studies recently, e.g., on their roles in performance of
information processing \cite{DQC,qip} and their relations to dynamics under
decoherence \cite{dynamics}. On the other hand, quantitative evaluation of
quantum discord involves an optimization procedure similar to that in the
EoF. The absence of a tractable method to tackle it makes comparison of
these two measures and relevant studies very obscure. Until now, the
explicit expression of quantum discord has only been obtained for certain
classes of two-qubit states \cite{luo,xstate}.

In this paper, we propose a scheme to evaluate the amount of quantum discord
and EoF via an intrinsic relation between the two quantities distributed in
the purification of given mixed states. The same relation of duality had
been revealed by Koashi and Winter in a context to explore the monogamy
nature of entanglement measures \cite{Koashi}. Here we demonstrate that the
elicited trilateral relationship can be applied to quantify quantum discord
and EoF and their ordering relation. In particularly, the scheme enables us
to fully characterize the discord and EoF for certain quantum states, e.g.,
an arbitrary two-qubit state reduced from pure three-qubit states and a
class of rank-$2$ mixed states of $4\times 2$ systems. Additionally, we
apply further the quantification scheme to describe the dynamical behavior
of quantum correlations for specified systems under decoherence.

\textit{Conditional entropy versus entanglement of formation in
purifications.} Let us start with the conditional entropy
\begin{equation}
S(\rho_B|A)\equiv \min_{\{A_k\}}\sum_kp_kS(\rho _B^k|A_k),  \label{condi}
\end{equation}
the central ingredient contained in the definition of the quantum discord.
We discriminate two different operations of $\{A_k\}$, the von Neumann
projective measurements $\{\Pi_k\equiv|k_A\rangle \langle k_A|;\,
k=1,\cdots,d\}$ and the generalized positive operator-valued measurements
(POVMs) $\{\mathcal{A}_k\}$ satisfying $\sum_k\mathcal{A}_k^{\dagger }
\mathcal{A}_k=I$. Their corresponding definitions of conditional entropy are
denoted by $S_I$ and $S_{II}$, respectively. Accordingly, we have two kinds
of definitions for quantum discord,
\begin{equation}
Q_{AB}^{I,II}=S(\rho_A)+S_{I,II}(\rho_B|A)-S(\rho_{AB}).  \label{q12}
\end{equation}

To proceed we invoke the so-called purification $|\Psi_{ABC}\rangle$ of $
\rho_{AB}$, whose partial trace on the ancillary $C$ gives rise to $\mathrm{
tr}_C|\Psi_{ABC}\rangle \langle \Psi_{ABC}|=\rho_{AB}$. Suppose that $C$ has
a dimension equal to the rank of $\rho _{AB}$, then all such purifications
should be equivalent up to local unitary transformations on $C$. With the
notion of $|\Psi _{ABC}\rangle $, each outcome of the orthogonal projective
measurement $\{|k_A\rangle \langle k_A|\}$ will be associated with a
relative state $|\psi_{BC}^k\rangle =\langle k_A|\Psi_{ABC}\rangle / \sqrt{
p_k}$, where $p_k$ is the probability of the $k$th outcome. The entropy of $B
$ (same as that of $C$), conditioned to the $k$th outcome, is precisely
captured by the entropy entanglement in $|\psi_{BC}^k\rangle$: $
S(\rho_B^k|A_k)=E(|\psi_{BC}^k\rangle )=S(\rho_C^k|A_k)$.
Furthermore, in view that the set of relative states $\{p_k,|\psi
_{BC}^k\rangle \}$ form actually an ensemble that realizes $\rho _{BC}$,
namely, $\rho _{BC}=\sum_kp_k|\psi _{BC}^k\rangle \langle \psi _{BC}^k|$, we
obtain
\begin{equation}
S_I(\rho_B|A)=\min_{\{\Pi_k\}} \sum_kp_kE(|\psi_{BC}^k\rangle )\equiv
E^{[d]}(\rho_{BC}),  \label{cond1}
\end{equation}
where $E^{[d]}(\rho_{BC})$ defines a $d$-component EoF, i.e., minimal
average entanglement of $\rho_{BC}$ over ensemble decompositions with only $d
$ components. In general there is $E^{[d]}(\rho_{BC})\geq E(\rho_{BC})$
since the minimization in the definition of $E(\rho _{BC})$ is taken over
all ensemble decompositions realizing $\rho _{BC}$ but the $d$-component
ensemble decompositions via the outcome of projective measurements $\{\Pi_k\}
$ on $A$ are only portion of them.

On the other hand, the outcome via the complete set of POVMs $\{\mathcal{A}
_k\}$ offers a distinct way to realize $\rho_{BC}$ in view that $\sum_k
\mathrm{tr}_A(\mathcal{A}_k|\Psi _{ABC}\rangle \langle \Psi_{ABC} |\mathcal{A
}_k^{\dagger})=\rho_{BC}$. Note that the ensemble generated here comprises
also those of mixed states. It follows from the concavity property of von
Neumann entropy that the minimum of the conditional entropy of Eq. (\ref
{condi}) is always reached with ensembles of pure states \cite{Koashi}.
Consequently, it happens that for the quantity $S_{II}(\rho_B|A)$ in which
POVMs on $A$ are promised, there exists
\begin{equation}
S_{II}(\rho _B|A)=E(\rho _{BC})=S_{II}(\rho _C|A).  \label{cond2}
\end{equation}
To make clear that a POVM realizing the optimal ensemble with minimal
average entanglement $E(\rho_{BC})$ can always be constructed, it is
instructive to invoke the following fact: by including an external system $E$
with an arbitrary high dimension and performing joint unitary evolutions on $
A$ and $E$, all ensembles of pure states reproducing $\rho _{BC}$ can be
generated via the outcome $\{p_{k^{\prime }k},|\psi _{BC}^{k^{\prime
}k}\rangle \}$ associated with von Neumann measurements $\{|k_A^{\prime
}k_E\rangle \langle k_A^{\prime }k_E|\}$ on $A$ and $E$, where
\begin{equation}
|\psi_{BC}^{k^{\prime }k}\rangle =\langle k_A^{\prime }k_E|U_{AE}
(|\Psi_{ABC}\rangle \otimes |0_E\rangle )/\sqrt{p_{k^{\prime }k}}.
\label{purif2}
\end{equation}
The action of a general POVM $\{\mathcal{A}_k\}$ on $|\Psi_{ABC}\rangle$
could be described in a similar way as
\begin{align}
\mathcal{A}_k|\Psi_{ABC}\rangle &= \langle k_E|U_{AE}(|\Psi_{ABC}\rangle
\otimes |0_E\rangle )  \nonumber \\
&= \sum_{k^{\prime }=1}^{d_A}\sqrt{p_{k^{\prime }k}}|k_A^{\prime }\rangle
|\psi_{BC}^{k^{\prime }k}\rangle .  \label{povm2}
\end{align}
To output the ensemble of pure states of Eq.\;(\ref{purif2}) through POVM
action, we revise the POVM (\ref{povm2}) by partitioning the operators
$\{\mathcal{A}_k\}$ into $\{\mathcal{A}_{k^{\prime },k}
=|k^{\prime}_A\rangle\langle  k^{\prime}_A|\mathcal{A}_k;\, k^{\prime
}=1,\cdots ,d\}$  with $d$ the dimension of $A$. It is then readily seen
that the new POVMs $\{\mathcal{A}_{k^{\prime },k}\}$ could give rise to the
general ensemble of relative states in Eq. (\ref{purif2}): $\mathcal{A}
_{k^{\prime },k}|\Psi_{ABC}\rangle  \rightarrow \{p_{k^{\prime }k}, |\psi
_{BC}^{k^{\prime }k}\rangle \}$.

The relation of (\ref{cond1}) and (\ref{cond2}) suggests a scheme to
quantify quantum discord and EoF (or the $d$-component EoF) through building
purifications of given mixed states. For example, in view that the EoF of
mixed states of two-qubit systems has already been perfectly resolved \cite
{wootters}, it indicates that the discord of all $n\times 2$ $(n\geq 2)$
density matrices with no more than two nonzero eigenvalues can be achieved
accordingly. Furthermore, by employing Eq. (\ref{cond2}) and the similar
relation for $S_{II}(\rho_A|C)$, one obtains
\begin{equation}
Q_{AB}^{II}-E(\rho_{AB})=Q_{AC}^{II}-Q_{CA}^{II}.  \label{amount}
\end{equation}
At this stage, it is clear that the ordering relation of the two quantities,
$Q_{AB}^{II}$ and $E(\rho_{AB})$, is essentially connected to the asymmetric
property of quantum discord, reflected via the lateral $\rho_{AC}$ (or $
\rho_{BC}$ if we consider $Q_{BA}^{II}$ in stead) in the purification. In
what follows we shall apply the scheme on some concrete quantum states so as
to evaluate the amount of their discord and EoF and the ordering relation.

\textit{Quantum discord and entanglement in pure states of three qubits.}
It is direct to apply the above scheme to solve the issue for arbitrary
two-qubit density matraces reduced from three-qubit pure states. In
view that a three-qubit pure state can be expressed generally as \cite
{minimum}
\begin{align}
|\psi _{ABC}\rangle & =\lambda _0|000\rangle +\lambda _1e^{i\varphi
}|010\rangle +\lambda _2|011\rangle   \nonumber \\
& \quad +\lambda _3|110\rangle +\lambda _4|111\rangle ,  \label{minimum}
\end{align}
the EoF of any pair could be worked out explicitly via the formula developed by
Wootters \cite{wootters},
\begin{equation}
E(\rho _{XY})=-x\log _2x-(1-x)\log _2(1-x),  \label{eof}
\end{equation}
where $x=\frac 12\left[ 1+\sqrt{1-C^2(\rho _{XY})}\right] $ and the
concurrence is obtained as
\begin{equation}
\begin{split}
C(\rho _{AB})& =2\lambda _0\lambda _3,\qquad C(\rho _{BC})=2\lambda
_0\lambda _2, \\
C(\rho _{AC})& =2|\lambda _2\lambda _3-\lambda _1\lambda _4e^{i\varphi }|.
\end{split}
\label{conc}
\end{equation}
Since the optimal ensemble decomposition realizing
the minimal average entanglement for each $\rho
_{XY}$ here has only two components, it can be achieved by projective
measurements on the third party $Z$. So there is $S_I(\rho
_X|Z)=S_{II}(\rho _X|Z)=E(\rho _{XY})$. As the conditional entropy has been
derived, the discord of each $\rho_{ZX}$ can be directly obtained according to
Eq.\ (\ref{q12}).

To illustrate the relation (\ref{amount}), we note that for the case
$\lambda _2=\lambda _3$, the state $\rho _{AC}$ is symmetric under
permutation hence $Q_{AC}=Q_{CA}$. Consequently, there is
\begin{equation}
Q_{AB}=E(\rho _{AB})=-\sum_{+,-}\frac{1\pm \Delta }2\log _2\frac{1\pm \Delta
}2,  \label{discord2}
\end{equation}
where $\Delta =\sqrt{1-\lambda _0^2\lambda _2^2}$\thinspace .

\textit{Quantum discord in} $4\times 2$\textit{\ systems with no more than
two nonzero eigenvalues.} The discord of any $n\times 2$ system $\rho _{AB}$
with only two nonzero eigenvalues can be recast to that of a $4\times 2$
system. This can be readily seen that for the purification $
|\psi_{ABC}\rangle $ of $\rho _{AB}$ the relative system $C$ could be of two
dimension, so that the reduced density matrix $\rho _{BC}$ (hence $\rho _A$)
has at most four nonzero eigenvalues. Since the EoF of two binary systems $
\rho_{BC}$ has perfectly resolved, it leads to the fact that the discord of
any such states can be explicitly obtained. To illustrate, we present below
an example of four-parameter family of rank-$2$ states formed as
\begin{equation}
\rho_{AB}=p_1|\psi_1\rangle \langle \psi_1| +p_2|\psi_2\rangle \langle
\psi_2|,  \label{state2}
\end{equation}
where
\begin{equation}
\begin{split}
|\psi_1\rangle &= \cos\phi |00\rangle +\sin\phi |11\rangle , \\
|\psi_2\rangle &= \sin\phi |a_30\rangle +\cos\phi |a_41\rangle,
\end{split}
\label{examp2a}
\end{equation}
are two normalized eigenvectors of $\rho _{AB}$ with
\begin{equation}
\begin{split}
|a_3\rangle &=\cos \theta _1|1\rangle +\sin \theta _1|2\rangle , \\
|a_4\rangle &=\cos \theta _2|0\rangle +\sin \theta _2|3\rangle .
\end{split}
\label{examp2}
\end{equation}
To calculate the quantum discord in Eq.\;(\ref{q12}), it is easy to obtain
that $S(\rho _{AB})=-\sum_{i=1}^2p_i\log p_i$ and $S(\rho_A)=-\sum_{i=1}^4
\lambda _i\log \lambda _i$, where
\begin{equation}
\begin{split}
\lambda_{1,2} &=\frac 12\sin^2\!\phi \Big( 1\pm \sqrt{1-4p_1p_2\sin\theta_1^2
}\; \Big), \\
\lambda_{3,4} &=\frac 12\cos^2\!\phi \Big(1\pm \sqrt{1-4p_1p_2\sin \theta
_2^2}\; \Big).
\end{split}
\label{lambda}
\end{equation}
The conditional entropy $S(\rho _B|A)$ can be derived via the equality of
Eq.\;(\ref{cond2}) and the purification herein could be expressed as $
|\psi_{ABC}\rangle =\sqrt{p_1}|\psi_1\rangle |0_C\rangle  +\sqrt{p_1}
|\psi_2\rangle |1_C\rangle$. 
Note that the optimal ensemble decomposition realizing $E(\rho_{BC})$ has
four components \cite{wootters}, this very ensemble can be achieved through
outcomes of the projective measurements on $A$ in view that $\dim A=4$. This
leads to the relation $S_I(\rho_B|A)=S_{II}(\rho_B|A)=E(\rho _{BC})$ for the
present system. Explicitly, its value is given by the formula of Eq.\;(\ref
{eof}) and the corresponding concurrence is now obtained as $%
C(\rho_{BC})=\max \{0,\, 2\lambda_m^c\!-\!\sum_{i=1}^4\lambda _i^c\}$, where
$\lambda_m^c$ is the largest of
\begin{equation}
\begin{split}
\lambda_{1,2}^c &=\frac{1}{2}\sin^2\!\phi \big[\sqrt{1-(p_1-p_2)^2} \pm 2
\sqrt{p_1p_2}\cos\theta_1 \big], \\
\lambda_{3,4}^c &=\frac{1}{2}\cos^2\!\phi \big[\sqrt{1-(p_1-p_2)^2} \pm 2
\sqrt{p_1p_2}\cos\theta_2 \big].
\end{split}
\label{lambdac}
\end{equation}

\textit{Deriving entanglement of formation via quantum discord}. The
quantitative calculation of EoF for mixed states is notoriously difficult
and the explicit expression is derived only for two-qubit systems \cite
{wootters} and very limited cases of high dimensional systems \cite{isotro}.
Noteworthily, the derivation of the relation of quantum correlations in
tripartite purifications suggests also a distinct way to calculate EoF in
virtue of the conditional entropy. In particular, let us consider the states
$\rho _{AB}$ of a $4\times 2$ system given by Eq.\;(\ref{state2}). By
examining the purification $|\psi _{ABC}\rangle $ it is seen that the
resulting state $\rho _{BC}$ is an X-class two-qubit state, with the nonzero
elements being
\begin{align}
\rho_{BC}^{11} &= p_1\cos^2\!\phi , \ \ \rho_{BC}^{14}=\rho_{BC}^{41} =\sqrt{
p_1p_2}\cos^2\!\phi \cos\theta_2,  \nonumber \\
\rho_{BC}^{22} &=p_2\sin^2\!\phi , \ \ \rho_{BC}^{23}=\rho_{BC}^{32} =\sqrt{
p_1p_2}\sin^2\!\phi \cos\theta_1,  \nonumber \\
\rho_{BC}^{33} &=p_1\sin^2\!\phi , \ \ \rho_{BC}^{44} = p_2\cos^2\!\phi .
\label{xstate}
\end{align}
In the two-qubit case, it has been proved \cite{kobes} that the projective
measurement is the optimal POVM to minimize the conditional entropy, and
evaluation of it for the X-class state has been resolved in Ref.\;
\onlinecite{xstate}. According to the scheme proposed previously, it means $
E(\rho_{AB})=E^{[2]}(\rho_{AB})$ and its expression can be explicitly
calculated. Particularly, for the case of $p_1=p_2=1/2$, the state $\rho
_{BC}$ reduces to a Bell-diagonal state (i.e., $\rho _B=\rho _C=I_2/2$). The
conditional entropy $S(\rho_B|C)$ in this case has a concise expression \cite
{luo} and so does the EoF of $\rho_{AB}$,
\begin{equation}
E(\rho_{AB})=S(\rho_B|C)=-\sum_{+,- }\frac{1\pm \chi }{2} \log_2 \frac{
1\pm\chi}{2},  \label{eof3}
\end{equation}
where $\chi =\max \{|\chi_1|,|\chi_2|,|\chi_3|\}$, with
\begin{equation}
\chi_1= -\cos 2\phi ,\ \ \chi_{2,3}=\cos^2\!\phi \cos\theta_2 \pm
\sin^2\!\phi \cos\theta_1.  \label{kapha}
\end{equation}

The calculation above actually offers a full characterization of the state
(\ref{state2}) for both its quantum discord and EoF. We plot in Fig.\;1 the
two quantities as functions of $\phi$, where $p_{1,2}=1/2$, $\theta_1=0$,
and $\theta_2=\pi/3$. The result shows that the EoF is lower than the
discord as $\sin^2\!\phi\lesssim0.070$ and $\sin^2\!\phi \gtrsim 0.711$, and
larger than the discord in the range $0.070<\sin^2\!\phi <0.711$.

\begin{figure}[tbp]
\begin{center}
\epsfig{figure=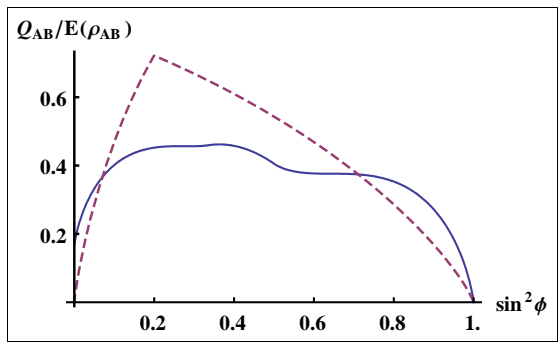,width=0.41\textwidth}
\end{center}
\caption{Quantum discord (solid line) versus EoF (dashed line) in the $
4\times 2$ system specified by Eq.\;(\ref{state2}), where the coefficients
are given by $p_{1,2}=1/2$, $\theta_1=0$, and $\theta_2=\pi/3$. Both
quantities are shown to be discrete functions of $k_1^2$, and the partition
of range is depicted as $\sin^2\!\phi\in (0,\frac{1}{3}), (\frac{1}{3},\frac{
1}{2}), (\frac{1}{2},1)$ for the discord, and $\sin^2\!\phi\in (0,\frac{1}{5}
), (\frac{1}{5},1)$ for EoF, respectively.}
\end{figure}

\begin{figure}[tbp]
\begin{center}
\epsfig{figure=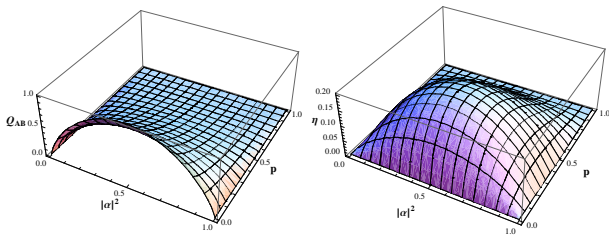,width=0.45\textwidth}
\end{center}
\caption{Dynamical behavior of quantum correlations under a phase-damping
condition [cf.~Eq.\;(\ref{photon})]. Plotted are quantum discord $Q_{AB}$
(left) and its difference from EoF, $\eta\equiv E(\rho_{AB})-Q_{AB}$
(right), as functions of $|\alpha|^2$ and $p=1-e^{-\gamma t}$.}
\end{figure}

As a final proposal of the paper we apply the derived results of
quantification on discord and EoF to characterize their dynamical behavior
in a typical physical process. Consider a two-qubit system initially
prepared in the state $|\psi_{AB}\rangle =\alpha |00\rangle +\beta |11\rangle
$, with $|\alpha|^2+|\beta|^2=1$. Suppose that one of the qubit is subjected
to a phase-damping environment (which can be realized, e.g., in optical
systems as one of the photon passes through a phase-damping channel \cite
{guo}). The system will then evolve as
\begin{align}
\rho_{AB} &= |\alpha |^2|00\rangle \langle 00| +|\beta |^2|11\rangle \langle
11|  \nonumber \\
&\quad +e^{-\gamma t} \big(\alpha \beta^{\ast}|00\rangle \langle 11|
+\alpha^{\ast}\beta |11\rangle\langle 00|\,\big),  \label{photon}
\end{align}
with the phase-damping rate $\gamma$. Since the state $\rho_{AB}$ has only
two nonzero eigenvalues, its quantum discord can be worked out explicitly
via the above derived results upon its purification $|\psi_{ABC}\rangle $ of
the three-qubit system. It turns out that $\rho_{BC}$ and $\rho_{AC}$
reduced from $|\psi_{ABC}\rangle$ are separable states. Therefore the
conditional entropy $S_{I,II}(\rho_B|A)=E(\rho_{BC})=0$. The quantum discord
of $\rho _{AB}$ is obtained simply as $Q_{AB}=S(\rho_A)-S(\rho_{AB})$, which
is fully specified by the spectrum of $\rho_A$ and $\rho_{AB}$, expressed in
detail as $\{|\alpha|^2,|\beta|^2\}$ and $\tfrac{1}{2}\pm \sqrt{\tfrac{1}{4}
-|\alpha\beta|^2(1-e^{-2\gamma t})}$, respectively.
We plot in Fig.\;2 the dynamical behavior of quantum discord $Q_{AB}$ and
its difference from EoF of $\rho_{AB}$, as function of $|\alpha|^2$ and $
p=1-e^{-\gamma t}$. The EoF is obtained from Eq.\;(\ref{eof}), in which the
concurrence of the present system is $C(\rho_{AB})=2|\alpha \beta |
e^{-\gamma t}$. It is shown that the EoF is always larger than discord in
the specified phase-damping process of this model.

In summary, two measures of quantum correlations, quantum discord and EoF,
have been investigated in unified purifications of given mixed quantum
states. We show that their amount and ordering relation can be evaluated via
the relationship of them distributing in different partners of the
purification. The scheme is then exploited to achieve analytical expressions
of quantum discord and EoF for some quantum states, including arbitrary
two-qubit density matrices reduced from pure three-qubit states and a class
of rank-$2$ mixed states of $4\times 2$ systems. In its applications to
physical systems, we show that the derived result of quantification on the
discord and EoF enables us to characterize their dynamical behavior in
certain typical physical processes. Finally, we mention other applications
proposed recently concerning the duality relation of quantum discord and
EoF, e.g., resolution of EoF for a class of Gaussian states \cite{Gaussian},
and the role of them in relation to the power of deterministic quantum
computation \cite{added}.

Support from the National Natural Science Foundation of China (10604043 and
10874254), and RGC Hong Kong (604709) is acknowledged.

\end{document}